\documentclass[preprint,aps,nofootinbib,showpacs,amsfonts,epsf]{revtex4}

\input{epsf.tex}

\newcommand{\E}{{\cal{E}}}
\newcommand{\s}{\sigma}

\renewcommand{\a}{\alpha}

\newcommand{\be}{\begin{equation}}
\newcommand{\ee}{\end{equation}}
\newcommand{\bea}{\begin{eqnarray}}
\newcommand{\eea}{\end{eqnarray}}
\newcommand{\ba}{\begin{array}}
\newcommand{\ea}{\end{array}}
\def\J#1#2#3#4{{#1} {\bf #2}, #3 (#4)}
\def\PRD{Phys. Rev. D}
\def\PR{Phys. Rev.}
\def\PRL{Phys. Rev. Lett.}
\def\PTP{Prog. Theor. Phys.}

\def\APN{Ann. Phys. (NY)}

\def\JMP{J. Math. Phys.}
\def\CPAM{Comm. Pure Appl. Math.}

\def\CQG{Class. Quantum Grav.}

\def\GRG{Gen. Relativ. Grav.}
\def\PLA{Phys. Lett. A}
\def\PLB{Phys. Lett. B}

\begin{document}
\draft
\title{Metric for two arbitrary Kerr sources}

\author{V.~S.~Manko$^\dagger$ and E.~Ruiz$^\ddagger$ }
\address{$^\dagger$Departamento de F\'\i sica, Centro de Investigaci\'on y
de Estudios Avanzados del IPN, A.P. 14-740, 07000 Ciudad de
M\'exico, Mexico\\$^\ddagger$Instituto Universitario de F\'{i}sica
Fundamental y Matem\'aticas, Universidad de Salamanca, 37008
Salamanca, Spain}

\begin{abstract}
The full metric describing a stationary axisymmetric system of two
arbitrary Kerr sources, black holes or hyperextreme objects,
located on the symmetry axis and kept apart in equilibrium by a
massless strut is presented in a concise explicit form involving
five physical parameters. The binary system composed of a
Schwarzschild black hole and a Kerr source is a special case not
covered by the general formulas, and we elaborate the metric for
this physically interesting configuration too.
\end{abstract}


\maketitle

\section{Introduction}

In our recent paper \cite{MRu} we have outlined an approach
towards elaborating a physically attractive description of two
arbitrary Kerr sources separated by a massless strut. We have
shown that obtaining such a description is a feasible task because
the axis condition -- the only one needed to be solved -- reduces
to an algebraic equation of the order not higher than quartic.
There we have also hinted that some clever trick might help
eventually obviate the resolution of the cumbersome quartic
equation and get concise expressions for the quantities $\s_1$ and
$\s_2$ related to the horizons of black holes in the subextreme
case. Writing that, we first of all had in mind our earlier work
\cite{MRu2} where we obtained a simple representation of the
well-known Kinnersley-Chitre metric \cite{KCh,Yam} and used it for
solving the axis condition, thus finding all possible binary
configurations of extreme Kerr black holes separated by a strut;
the use of the NUT parameter \cite{NTU} $J_0$ in that analysis was
a key ingredient for choosing a convenient parametrization by
solving the asymptotic flatness condition $J_0=0$, and so it
looked to us plausible to explore the expression of $J_0$ in the
more general situation involving non-extreme Kerr constituents in
order to obtain a second constraint which, together with the axis
condition, would provide a system of two algebraic equations for
finding the aforementioned quantities $\s_1$ and $\s_2$ uniquely.

It appears that the above idea has been successfully accomplished
recently by Cabrera-Munguia \cite{Cab} who, being well familiar
with the paper \cite{MRu2}, has succeeded in obtaining the desired
expressions for $\s_1$ and $\s_2$ through a straightforward
algebra involving the resolution of a quadratic equation, thus
making an important step forward towards a definitive description
of binary configurations of Kerr sources. At the same time,
despite the author's claim that the metric given by him covers
both the black-hole and hyperextreme sectors of the double-Kerr
solution, it can be shown that this is not true, and consequently,
like in the author's earlier paper on two identical Kerr sources
\cite{CCL}, his formulas have conceptual problems that make them
inappropriate for the description of hyperextreme Kerr objects.
Such a situation is obviously undesirable. Indeed, first of all,
since both the black holes and naked singularities are able to
form configurations with struts, it is of course likely to have
solutions able to describe both types of the Kerr sources. But
more importantly, provided that some binary configurations with
struts may contain configurations without struts as particular
limiting cases corresponding to vanishing interaction force, the
former configurations must be reducible to the latter ones, the
majority of which are actually configurations involving
hyperextreme Kerr constituents \cite{DHo,MRS}. It might also be
noted that in view of the discovery of the black hole-naked
singularity dualism \cite{MRu3}, the particular binary systems
representing this phenomenon cannot be correctly treated by the
solutions restricted exclusively to the black-hole sector.

The objective of the present paper is to combine together our
previous results on the binary configurations of Kerr sources with
the recent expressions for $\s_1$ and $\s_2$ found in \cite{Cab}
for obtaining a concise physical representation of the metric for
two arbitrary Kerr sources separated by a massless strut which
would describe in a unified manner both the black-hole and
hyperextreme constituents. We shall also work out a particular
4-parameter metric describing a physically interesting
`Schwarzschild-Kerr' configuration which is not covered by the
general formulas and for which only an equilibrium solution
without a strut was obtained earlier in the literature
\cite{MRu4}; it will illustrate well the transformation of a
black-hole binary system into a configuration involving a
hyperextreme Kerr source within the same extended solution.

\section{The solution in physical parameters}

As it follows from the papers \cite{MRu,Cab}, the Ernst complex
potential $\E$ \cite{Ern} of the exact solution for two aligned
Kerr sources is defined on the symmetry axis by the expression
\be e(z)=\frac{z^2-(M+ia)z+s-\mu+i(\tau+\delta)}
{z^2+(M-ia)z+s+\mu+i(\tau-\delta)}, \label{axisG} \ee
where $M$ is the total mass of the binary system, $a$ the
rotational parameter, and the real quantities $s$, $\mu$, $\tau$,
$\delta$ are related to the individual physical characteristics of
the sources in the following way:
\bea s&=&-\frac{1}{4}[R^2+2(\s_1^2+\s_2^2-M^2+a^2)], \quad
\delta=Ma-m_1a_1-m_2a_2, \nonumber\\
\tau&=&\frac{1}{2}R(a_2-a_1)+
\frac{(R+M)[m_2a_1(R+2m_1)-m_1a_2(R+2m_2)]}{(R+M)^2+a^2}, \nonumber\\
\mu&=&\frac{1}{2M}[R(\s_1^2-\s_2^2)-2a\tau]. \label{dk} \eea
The set of {\it five} arbitrary real parameters is
$\{m_1,m_2,a_1,a_2,R\}$, $m_1$ and $m_2$ being the individual
Komar \cite{Kom} masses of the Kerr constituents, $a_1=j_1/m_1$
and $a_2=j_2/m_2$ their individual Komar angular momenta per unit
mass, and $R$ the coordinate distance between the centers of the
sources. The total mass $M$ and total angular momentum $J$ of the
system therefore have the form
\be M=m_1+m_2, \quad J=m_1a_1+m_2a_2, \label{MJ} \ee
while $a$ satisfies the cubic equation
\be \frac{(a_1+a_2-a)(R^2-M^2+a^2)}{2(R+M)}-Ma+J=0. \label{AM} \ee
The remaining quantities $\s_1$ and $\s_2$, which represent the
half-lengths of the horizons of black holes in the subextreme
case, are given by the formulas \cite{Cab}
\bea \s_1=\sqrt{m_1^2-a_1^2+4m_2a_1 \frac{
[m_1(a_1-a_2+a)+Ra][(R+M)^2+a^2]+m_2a_1a^2}
{[(R+M)^2+a^2]^2}}, \nonumber\\
\s_2=\sqrt{m_2^2-a_2^2+4m_1a_2 \frac{
[m_2(a_2-a_1+a)+Ra][(R+M)^2+a^2] +m_1a_2a^2}{[(R+M)^2+a^2]^2}},
\label{s1s2} \eea
and it is clear that $\s_1$ and $\s_2$ can take on real or pure
imaginary values.

With the axis data (\ref{axisG}) thus defined, the equation
\be e(z)+\bar e(z)=0, \label{Sibga} \ee
(a bar over a symbol means complex conjugation) has four roots
$\a_n$, $n=1,2,3,4$, of the form
\be \a_1=\frac{1}{2}R+\s_1, \quad \a_2=\frac{1}{2}R-\s_1, \quad
\a_3=-\frac{1}{2}R+\s_2, \quad \a_4=-\frac{1}{2}R-\s_2, \label{an}
\ee
and the above $\a_n$ define the location of the sources on the
symmetry axis, a pair of real-valued $\a$'s determining a black
hole, and a pair of complex conjugate $\a$'s determining a
hyperextreme object. Fig.~1 illustrates that there are three
generic types of binary configurations of non-extreme Kerr
constituents: the ``subextreme-subextreme'',
``subextreme-hyperextreme'' and ``hyperextreme-hyperextreme'',
among which only the first one is entirely composed of black
holes.

The expression of the Ernst complex potential $\E$ in the whole
$(\rho,z)$ space is obtainable from (\ref{axisG}) by means of
Sibgatullin's integral method \cite{Sib} (for details the reader
is referred to \cite{MRu} and references therein), and it reads as
follows
\bea \E&=&(A-B)/(A+B),
\nonumber\\
A&=&[R^2-(\s_1+\s_2)^2](R_+-R_-)(r_+-r_-)
-4\s_1\s_2(R_+-r_-)(R_--r_+), \nonumber\\
B&=&2\s_1(R^2-\s_1^2+\s_2^2)(R_--R_+)
+2\s_2(R^2+\s_1^2-\s_2^2)(r_--r_+) \nonumber\\
&&+4R\s_1\s_2(R_++R_--r_+-r_-), \label{EG} \eea
where the functions $R_\pm$ and $r_\pm$ have been found to be
defined by the formulas
\bea r_\pm&=&\mu_0^{-1}\frac{(\pm\s_1-m_1-ia_1)[(R+M)^2+a^2]
+2a_1[m_1a+iM(R+M)]} {(\pm\s_1-m_1+ia_1)[(R+M)^2+a^2]
+2a_1[m_1a-iM(R+M)]}\,\tilde r_\pm, \nonumber\\
R_\pm&=&-\mu_0\,\frac{(\pm\s_2+m_2-ia_2)[(R+M)^2+a^2]
-2a_2[m_2a-iM(R+M)]} {(\pm\s_2+m_2+ia_2)[(R+M)^2+a^2]
-2a_2[m_2a+iM(R+M)]}\,\tilde R_\pm, \label{RrG} \\
\tilde r_\pm&=&\sqrt{\rho^2+\left(z-\frac{1}{2}R\pm\s_1\right)^2},
\quad \tilde R_\pm=\sqrt{\rho^2+
\left(z+\frac{1}{2}R\pm\s_2\right)^2}, \quad
\mu_0:=\frac{R+M-ia}{R+M+ia}. \nonumber \eea

We have checked that the potential (\ref{EG}) satisfies
identically the Ernst equation~\cite{Ern}, namely,
\be (\E+\bar\E)\Delta\E =2(\nabla\E)^2, \label{EE} \ee
for any type of the Kerr constituents in the binary configuration.

The metric functions $f$, $\gamma$ and $\omega$ of the solution
for two Kerr sources, entering the line element
\be d s^2=f^{-1}[e^{2\gamma}(d\rho^2+d z^2)+\rho^2 d\varphi^2]-f(d
t-\omega d\varphi)^2, \label{Papa} \ee
have the form \cite{MRu}
\bea f&=&\frac{A\bar A-B\bar B}{(A+B)(\bar A+\bar B)}, \quad
e^{2\gamma}=\frac{A\bar A-B\bar B}{16|\s_1|^2|\s_2|^2K_0^2 \tilde
R_+\tilde R_-\tilde r_+\tilde r_-}, \quad \omega=2a-\frac{2{\rm
Im}[G(\bar A+\bar
B)]}{A\bar A-B\bar B}, \nonumber\\
G&=&-zB +\s_1(R^2-\s_1^2+\s_2^2)(R_--R_+)(r_++r_-+R) \nonumber\\
&&+\s_2(R^2+\s_1^2-\s_2^2)(r_--r_+)(R_++R_--R) \nonumber\\
&&-2\s_1\s_2\{2R[r_+r_--R_+R_--\s_1(r_--r_+)+\s_2(R_--R_+)]
\nonumber\\ &&+(\s_1^2-\s_2^2)(r_++r_--R_+-R_-)\}, \label{mfG}
\eea
where for $K_0$ we have got the following very simple formula
\be K_0=\frac{[(R+M)^2+a^2] [R^2-(m_1-m_2)^2+a^2]-4m_1^2m_2^2a^2}
{m_1m_2[(R+M)^2+a^2]}. \label{K0} \ee

Therefore, we have given, in terms of five physical parameters, a
complete description of the Ernst potential and entire metric for
two arbitrary Kerr sources separated by a massless strut (by
construction, the axis condition is satisfied automatically). The
metric is asymptotically flat and is valid for all three types of
binary configurations involving subextreme and/or hyperextreme
Kerr constituents, and we have confirmed this with a computer
check.

At this point, it is worth noting that the expressions for the
Ernst potential and metric functions given in the papers
\cite{Cab,CCL} are not valid in the hyperextreme case. One may
come to this conclusion by observing that for instance the
quantities $\bar s_\pm/s_\pm$ introduced in \cite{CCL} as unitary
complex objects for all types of binary systems, should not be
unitary when these describe hyperextreme constituents (see, e.g.,
Ref.~\cite{MRu} for the definition and properties of the objects
$X_i$).

Since the metric presented in \cite{Cab} is plagued with
conceptual errors and does not satisfy the field equations when
one of the constituents is hyperextreme, it is necessary to verify
whether some formulas derived in \cite{Cab} for the subextreme
case are also applicable to the configurations with hyperextreme
sources. This first of all refers to the expression of the
interaction force ${\cal F}=(e^{-\gamma_0}-1)/4$ \cite{Isr,Wei},
$\gamma_0$ being the constant value of the metric function
$\gamma$ on the strut, which in the black-hole case was shown to
have the form
\be {\cal F}=
\frac{m_1m_2[(R+M)^2-a^2]}{(R^2-M^2+a^2)[(R+M)^2+a^2]}, \label{F}
\ee
but for which a formal use of the formulas of the paper \cite{Cab}
would yield a different expression in the hyperextreme case.
Fortunately, the formulas of the present paper reveal that the
above formula (\ref{F}) still holds when the Kerr sources are
hyperextreme. This conclusion is important for performing a
correct transition to the case of equilibrium binary
configurations without struts \cite{MRu5} when $a=\pm(R+M)$. Let
us also mention that vanishing of ${\cal F}$ when $m_1$ (or $m_2$)
is equal to zero means the reduction of the two-body metric to a
single Kerr solution \cite{Ker} independently of the value of
$a_1$ (or $a_2$).

Moreover, when analyzing the thermodynamical properties of the
black-hole constituent in a ``subextreme-hyperextreme'' binary
configuration, one has to be sure that the formulas for the
horizon area $A_i$, surface gravity $\kappa_i$ and horizon's
angular velocity $\Omega^H_i$ of $i$th black-hole constituent,
$i=1,2$, derived in \cite{Cab} with the aid of Tomimatsu's
formulas \cite{Tom}, namely,
\bea \frac{A_1}{4\pi}&=&\frac{\sigma_1}{\kappa_1}=\frac{
\{(m_1+\s_1)[(R+M)^2+a^2]-2m_1a_1a\}^2 +a_1^2(R^2-M^2+a^2)^2}
{[(R+M)^2+a^2][(R+\s_1)^2-\s_2^2]}, \nonumber\\
\frac{A_2}{4\pi}&=&\frac{\sigma_2}{\kappa_2}=\frac{
\{(m_2+\s_2)[(R+M)^2+a^2]-2m_2a_2a\}^2 +a_2^2(R^2-M^2+a^2)^2}
{[(R+M)^2+a^2][(R+\s_2)^2-\s_1^2]}, \nonumber\\
\Omega^H_1&=&\frac{m_1-\s_1}{2m_1a_1}, \quad
\Omega^H_2=\frac{m_2-\s_2}{2m_2a_2}, \label{A12} \eea
and satisfying the Smarr mass formula \cite{Sma}
\be m_i=\frac{1}{4\pi}\kappa_iA_i+2\Omega^H_ij_i
=\s_i+2\Omega^H_ij_i, \quad i=1,2, \label{Smar} \ee
are also valid for a black hole in the presence of a hyperextreme
object. A direct check shows after some endeavor that this is
indeed the case.

We now turn to consideration of a nontrivial special case not
covered by the general formulas.

\section{The `Schwarzschild-Kerr' binary configuration}

The formulas presented in the previous section provide one with a
powerful tool for studying the behavior and physical properties of
two interacting arbitrary Kerr sources with almost the same ease
as the known single black hole spacetimes. For the analysis of a
particular binary configuration one only needs to assign concrete
values to the physical parameters $m_1$, $m_2$, $a_1$, $a_2$ and
$R$, then find the corresponding $a$ from the cubic equation
(\ref{AM}), the quantities $\s_1$ and $\s_2$ from (\ref{s1s2}),
and determine the functions $r_\pm$ and $R_\pm$ with the aid of
formulas (\ref{RrG}); these being substituted into (\ref{EG}) give
the form of the Ernst potential, while the substitutions into
(\ref{mfG}) and (\ref{K0}) yield the metric of that particular
binary configuration. However, one cannot assign zero value to
$a_1$ or to $a_2$ in the general formulas (\ref{RrG}) because a
subtle degeneration then occurs and these formulas fail to
describe the binary configuration correctly. This explains in
particular why the case of a system comprised of a rotating and a
non-rotating sources evaded the researchers for so many years like
a real ghost, and for example was not extracted from the general
expressions in the paper \cite{DHo}. Note that by setting, say,
$a_1=0$ in (\ref{RrG}) we get $r_\pm/\tilde r_\pm=\mu_0^{-1}$ and
the solution becomes problematic. Therefore, this special case
needs a separate analysis.

As a preliminary, let us first note that for some applications
with yet unknown precise values of the masses and angular momenta
of the constituents the resolution of the cubic equation
(\ref{AM}) can be circumvented after introducing the constant $a$
as arbitrary parameter instead of $a_1$ or $a_2$. Indeed, equation
(\ref{AM}) is linear in $a_i$ and hence can be trivially solved,
say, for $a_2$, thus changing the set of arbitrary parameters
$\{m_1,m_2,a_1,a_2,R\}$ into a new set $\{m_1,m_2,a_1,a,R\}$, the
physical characteristic $a_2$ being related to the parameters of
the latter set in a simple way by (\ref{AM}). Such a redefinition
of the parameters, as will be seen below, allows to describe
correctly and in a concise form the desired long-searched-for
binary system  composed of a Schwarzschild black hole and a Kerr
source separated by a massless strut.

The `Schwarzschild-Kerr' configuration is a 4-parameter
specialization of the general case when one of the angular momenta
is set equal to zero. By choosing $a_1=0$, we convert the upper
constituent into a Schwarzschild black hole, while the lower
constituent is a rotating Kerr source that could be a black hole
(real $\s_2$) or a naked singularity (pure imaginary $\s_2$). It
follows trivially from (\ref{s1s2}) that vanishing of $a_1$
implies $\s_1=m_1$, which means that the upper constituent can
never be a naked singularity, no matter how large is the angular
momentum per unit mass $a_2$ of the lower constituent. Note that
although the condition $a_1=0$ causes drastic simplification of
the expression for $\s_1$, the use of this condition in the
expression for $\s_2$ does not lead at first glance to a
considerable change in the aspect of that quantity and actually
does not simplify much the resolution of the cubic equation
(\ref{AM}) for $a$. However, as we have already mentioned, the
latter equation can be easily solved for $a_2$, yielding
\be a_1=0, \quad a_2=\frac{a[(R+M)^2+a^2]}{(R+m_2)^2-m_1^2+a^2},
\label{a2} \ee
which permits us to introduce $a$ as arbitrary parameter instead
of $a_2$ in all the formulas and simplify considerably the form of
$\s_2$, getting
\be \s_2=\sqrt{m_2^2-a^2+2a^2\rho_0}, \quad \rho_0:=\frac{2m_1m_2}
{(R+m_2)^2-m_1^2+a^2}, \label{s2} \ee
as well as the form of $s$, $\delta$, $\tau$ and $\mu$ in the axis
data (\ref{axisG}):
\bea s&=&\frac{\rho_0}{2}[(R+m_2)^2-m_1^2-a^2] -\frac{R^2}{4},
\quad
\delta=\frac{\rho_0a(R^2-M^2+a^2)}{2m_2}, \nonumber\\
\tau&=&\frac{1}{2}Ra- \rho_0a(R+M), \quad
\mu=\frac{1}{2}R(m_1-m_2) +\rho_0a^2. \label{sdtm} \eea
Then the substitution of (\ref{sdtm}) into the formula (72) of
\cite{MRu} leads to the correct expressions for $r_\pm$ and
$R_\pm$,
\bea r_\pm&=&\mp\frac{R\mp m_1+m_2+ia} {R\mp m_1+m_2-ia}\,\tilde
r_\pm, \quad R_\pm=\frac{\mp\s_2+ia(1-\rho_0)}
{m_2-ia\rho_0}\,\tilde
R_\pm, \nonumber\\
\tilde r_\pm&=&\sqrt{\rho^2+\left(z-\frac{1}{2}R\pm m_1\right)^2},
\quad \tilde
R_\pm=\sqrt{\rho^2+\left(z+\frac{1}{2}R\pm\s_2\right)^2},
\label{RrP} \eea
thus resolving the problem of the limit $a_1=0$ in the general
expressions (\ref{RrG}) of the previous section.

The Ernst potential and all metric functions of the
`Schwarzschild-Kerr' binary configuration, accounting for
(\ref{EG}) and (\ref{mfG}), assume the following final form
\bea \E&=&\frac{A-B}{A+B}, \quad f=\frac{A\bar A-B\bar
B}{(A+B)(\bar A+\bar B)}, \quad e^{2\gamma}=\frac{A\bar A-B\bar
B}{16|\s_2|^2K_0^2\tilde
R_+\tilde R_-\tilde r_+\tilde r_-}, \nonumber\\
\omega&=&2a-\frac{2{\rm
Im}[G(\bar A+\bar B)]}{A\bar A-B\bar B}, \nonumber\\
A&=&[R^2-(m_1+\s_2)^2](R_+-R_-)(r_+-r_-)
-4m_1\s_2(R_+-r_-)(R_--r_+), \nonumber\\
B&=&2m_1(R^2-m_1^2+\s_2^2)(R_--R_+)
+2\s_2(R^2+m_1^2-\s_2^2)(r_--r_+) \nonumber\\
&&+4Rm_1\s_2(R_++R_--r_+-r_-), \nonumber\\
G&=&-zB +m_1(R^2-m_1^2+\s_2^2)(R_--R_+)(r_++r_-+R) \nonumber\\
&&+\s_2(R^2+m_1^2-\s_2^2)(r_--r_+)(R_++R_--R) \nonumber\\
&&-2m_1\s_2\{2R[r_+r_--R_+R_--m_1(r_--r_+)+\s_2(R_--R_+)]
\nonumber\\ &&+(m_1^2-\s_2^2)(r_++r_--R_+-R_-)\}, \label{mf_SK}
\eea
with
\be K_0=\frac{[(R+m_1)^2-\s_2^2] [(R+m_2)^2-m_1^2+a^2]}
{m_2[(R+M)^2+a^2]}. \label{K0P} \ee

It is worthwhile noting that the static limit $a=0$ in the above
metric does not represent any difficulty and leads to the
Bach-Weyl solution for two Schwarzschild black holes \cite{BWe}.

A peculiar feature of the `Schwarzschild-Kerr' configuration is
that, whereas the upper constituent in it is always a black hole,
the lower constituent can be either a black hole or a naked
singularity, depending on the values of the rotational parameter
$a$. When both constituents are subextreme, the areas of their
horizons assume the form
\bea A_1&=&\frac{16\pi m_1^2[(R+M)^2+a^2]}
{(R+m_1)^2-\s_2^2}, \nonumber\\
A_2&=&\frac{8\pi m_2[(R+M)^2+a^2]\{(m_2+\s_2)
[(R+m_1)^2-m_2^2+a^2] -2a^2\rho_0(R+M)\}}
{[(R+m_2)^2-m_1^2+a^2][(R+m_1)^2-\s_2^2]}, \label{A12P} \eea
and one can see that these are substantially affected by the black
hole interaction. At the same time, in the case of balance
($a=\pm(R+M)$), when gravitational attraction is equal to
spin-spin repulsion, the formula for $A_2$ is no longer valid
because $\s_2$ then becomes a pure imaginary quantity \cite{MRu4}
\be \s_2=i\sqrt{R^2-m_1^2+2R[M+m_1^2(R+m_2)^{-1}]}, \label{s2I}
\ee
which means that in the equilibrium state the lower constituent is
a naked singularity. Nevertheless, the expression for $A_1$ still
holds and coincides with the one obtained in \cite{MRu4}:
$A_1=16\pi m_1^2(1+m_2/R)$. The solution (\ref{mf_SK}) describes
correctly both sectors of the `Schwarzschild-Kerr' configuration,
thus giving unified treatment of the subextreme and hyperextreme
sources as inseparable ingredients of the same global spacetime.

\section{Conclusion}

The construction of the solution for two arbitrary Kerr sources
separated by a massless strut in a completely physical
parametrization may be considered as an important new contribution
into the list of physically meaningful spacetimes of Einstein's
general relativity. To a certain extent, the 5-parameter metric
presented in this paper draws a line beneath a long period of
extensive studies of the famous double-Kerr solution that had
started immediately after its discovery by Kramer and Neugebauer
\cite{KNe} almost four decades ago. Although many physical
properties of the black-hole and hyperextreme sectors of the
double-Kerr spacetime were clarified in the 80s of the last
century \cite{OSa,TKi,Tom2,Yam,Hoe,DHo}, it was eventually the
extended version of the double-Kerr solution \cite{MRu6,MRS}
obtained within the framework of Sibgatullin's integral method
\cite{Sib} that made possible the unified treatment of the sub-
and hyperextreme Kerr constituents in binary configurations and
the introduction of the physical parametrizations on the basis of
the standard parameters of the extended solution. After obtaining
a physical representation of the double-Kerr equilibrium problem
\cite{MRu5}, we were convinced that a more general binary
configuration of arbitrary Kerr sources separated by a massless
strut can be also rewritten in terms of physical parameters, and
we are glad that our expectations have finally come true.

\section*{Acknowledgments}

We dedicate this paper to the memory of Nail Sibgatullin who would
be 75 this year. Financial support from CONACYT of Mexico and from
MINECO/FEDER of Spain through the Project FIS2015-65140-P is
gratefully acknowledged.


\begin{figure}[htb]
\centerline{\epsfysize=120mm\epsffile{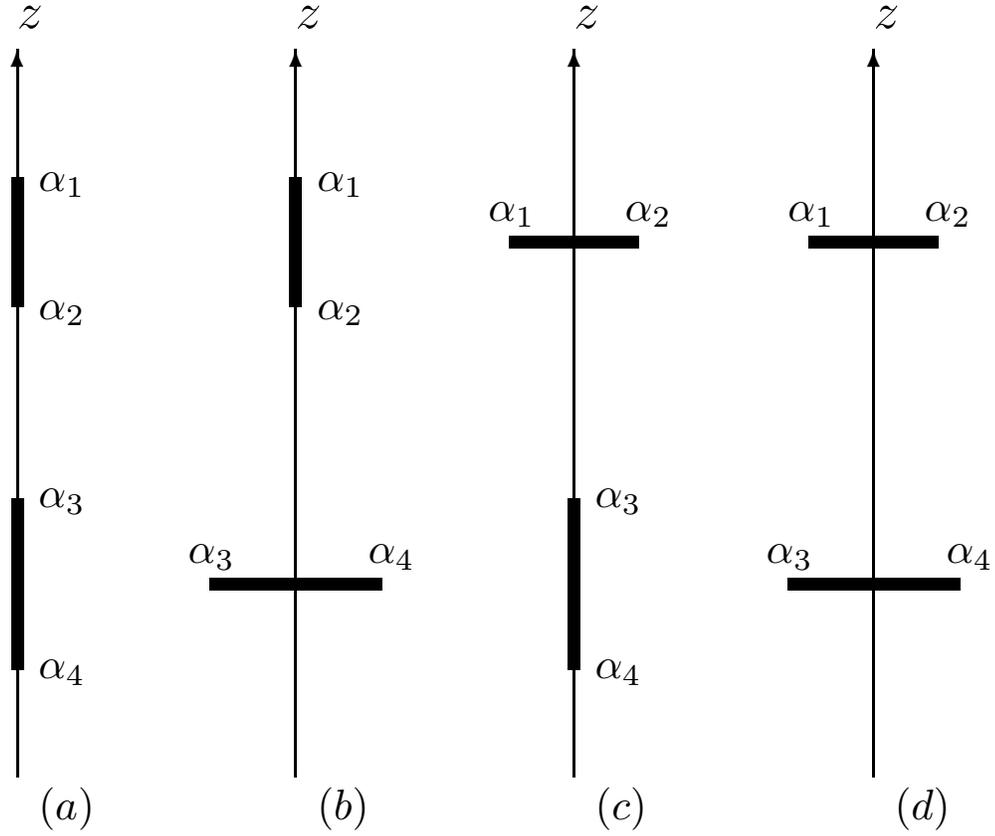}} \caption{Three
types of binary configurations composed of nonextreme Kerr
sources: ($a$) `sub\-extreme-subextreme', ($b$)-($c$)
`subextreme-hyperextreme', and ($d$) `hyperextreme-hyperextreme'.}
\end{figure}


\begin{references}

\bibitem{MRu} V.~S.~Manko and E.~Ruiz, \J{\PRD}{96}{104016}{2017}.

\bibitem{MRu2} V. S. Manko and E. Ruiz, \J{\PTP}{125}{1241}{2011}.

\bibitem{KCh} W. Kinnersley and D. M. Chitre, \J{\JMP}{19}{2037}{1978}.

\bibitem{Yam} M. Yamazaki, \J{\PTP}{63}{1950}{1980}.

\bibitem{NTU} E. T. Newman, L. A. Tamburino, and T. Unti,
\J{\JMP}{4}{915}{1963}.

\bibitem{Cab} I. Cabrera-Munguia, arXiv:1806.0544v.1

\bibitem{CCL} I. Cabrera-Munguia, V. E. Ceron, L. A. L\'opez,
and O. Pedraza, \J{\PLB}{10}{772}{2017}.

\bibitem{DHo} W. Dietz and C. Hoenselaers, \J{\APN}{165}{319}{1985}.

\bibitem{MRS} V. S. Manko, E. Ruiz and J. D. Sanabria-G\'omez,
\J{\CQG}{17}{3881}{2000}.

\bibitem{MRu3} V. S. Manko and E. Ruiz, arXiv:1803.03301.

\bibitem{MRu4} V.~S.~Manko and E.~Ruiz, \J{\GRG}{44}{2891}{2012}.

\bibitem{Ern} F.~J.~Ernst, \J{\PR}{167}{1175}{1968}.

\bibitem{Kom} A.~Komar, \J{\PR}{113}{934}{1959}.

\bibitem{Sib} N. R. Sibgatullin, Oscillations and Waves in Strong
Gravitational and Electromagnetic Fields (Berlin: Springer, 1991);
V.~S.~Manko and N.~R.~Sibgatullin, \J{\CQG}{10}{1383}{1993}.

\bibitem{Isr} W. Israel, \J{\PRD}{15}{935}{1977}.

\bibitem{Wei} G.~Weinstein, \J{\CPAM}{43}{903}{1990}.

\bibitem{MRu5} V. S. Manko and E. Ruiz, \J{\PRD}{92}{104004}{2015}.

\bibitem{Ker} R. P. Kerr, \J{\PRL}{11}{237}{1963}.

\bibitem{Tom} A. Tomimatsu, \J{\PTP}{72}{73}{1984}.

\bibitem{Sma} L.~Smarr, \J{\PRL}{30}{71}{1973}.

\bibitem{BWe} R. Bach and and H.~Weyl, \J{Math. Z.}{13}{134}{1922}.

\bibitem{KNe} D. Kramer and G. Neugebauer, \J{\PLA}{75}{259}{1980}.

\bibitem{OSa} K. Oohara and H. Sato, \J{\PTP}{65}{1891}{1981}.

\bibitem{TKi} A. Tomimatsu and M. Kihara, \J{\PTP}{67}{1406}{1982}.

\bibitem{Tom2} A. Tomimatsu, \J{\PTP}{70}{385}{1983}.

\bibitem{Yam} M. Yamazaki, \J{\PRL}{50}{1027}{1983}.

\bibitem{Hoe} C. Hoenselaers, \J{\PTP}{72}{761}{1984}.

\bibitem{MRu6} V.~S.~Manko and E.~Ruiz, \J{\CQG}{15}{2007}{1998}.

\end{references}
\end{document}